\documentclass[10pt,conference, doublecolumn]{IEEEtran}

\IEEEoverridecommandlockouts
\usepackage{cite}
\usepackage{amsmath,amssymb,amsfonts}
\usepackage{algorithmic}
\usepackage{graphicx}
\usepackage{float}   
\usepackage{textcomp}
\usepackage{epstopdf}
\usepackage[font=small,skip=0pt]{caption}
\usepackage{multirow}
\usepackage[utf8]{inputenc}
\usepackage[english]{babel}
\usepackage[usenames, dvipsnames]{color}

\def\BibTeX{{\rm B\kern-.05em{\sc i\kern-.025em b}\kern-.08em
    T\kern-.1667em\lower.7ex\hbox{E}\kern-.125emX}}

\usepackage{caption}
\usepackage{subcaption}

\usepackage[utf8]{inputenc}
\usepackage[english]{babel}

\usepackage{scalerel}
\usepackage{tikz}
\usetikzlibrary{svg.path}

\definecolor{orcidlogocol}{HTML}{A6CE39}
\tikzset{
  orcidlogo/.pic={
    \fill[orcidlogocol] svg{M256,128c0,70.7-57.3,128-128,128C57.3,256,0,198.7,0,128C0,57.3,57.3,0,128,0C198.7,0,256,57.3,256,128z};
    \fill[white] svg{M86.3,186.2H70.9V79.1h15.4v48.4V186.2z}
                 svg{M108.9,79.1h41.6c39.6,0,57,28.3,57,53.6c0,27.5-21.5,53.6-56.8,53.6h-41.8V79.1z M124.3,172.4h24.5c34.9,0,42.9-26.5,42.9-39.7c0-21.5-13.7-39.7-43.7-39.7h-23.7V172.4z}
                 svg{M88.7,56.8c0,5.5-4.5,10.1-10.1,10.1c-5.6,0-10.1-4.6-10.1-10.1c0-5.6,4.5-10.1,10.1-10.1C84.2,46.7,88.7,51.3,88.7,56.8z};
  }
}

\newcommand\orcidicon[1]{\href{https://orcid.org/#1}{\mbox{\scalerel*{
\begin{tikzpicture}[yscale=-1,transform shape]
\pic{orcidlogo};
\end{tikzpicture}
}{|}}}}

\usepackage{hyperref} 

\def\BibTeX{{\rm B\kern-.05em{\sc i\kern-.025em b}\kern-.08em
    T\kern-.1667em\lower.7ex\hbox{E}\kern-.125emX}}
\AtBeginDocument{\definecolor{ojcolor}{cmyk}{0.93,0.59,0.15,0.02}}

\begin{document}

\title{Featureless Wireless Communications using Enhanced Autoencoder}

\author{
\IEEEauthorblockN{Ruhui Zhang, Wei Lin, Binbin Chen} 

\thanks{Ruhui Zhang is with the Institute of Advanced Study in Mathematics, Harbin Institute of Technology, Harbin, China. (Email:
ruhui\_zhang@hit.edu.cn)}  
\thanks{Wei Lin and Binbin Chen are with the Pillar of Information Systems Technology and Design, Singapore University of Technology and Design, Singapore. (Email: wei\_lin@mymail.sutd.edu.sg, binbin\_chen@sutd.edu.sg) }
}


\maketitle
\vspace{-0.0cm}
\begin{abstract}
Artificial intelligence (AI) techniques—particularly autoencoders (AEs)—have gained significant attention in wireless communication systems. This paper investigates using an AE to generate featureless signals with a low probability of detection and interception (LPD/LPI). Firstly, we introduce a novel loss function that adds a Kullback–Leibler (KL) divergence term to the categorical cross-entropy, enhancing the noise-like characteristics of AE-generated signals while preserving block error rate (BLER). Secondly, to support long source message blocks for the AE's inputs, we replace one-hot inputs of source blocks with binary inputs pre-encoded by conventional error-correction coding schemes. The AE’s outputs are then decoded back to the source blocks using the same scheme. This design enables the AE to learn the coding structure, yielding superior BLER performance on coded blocks and the BLER of the source blocks is further decreased by the error-correction decoder. Moreover, we also validate the AE-based communication system in the over-the-air communication. Experimental results demonstrate that our proposed methods improve the featureless properties of AE signals and significantly reduce the BLER of message blocks, underscoring the promise of our AE-based approach for secure and reliable wireless communication systems.


\end{abstract}


\section{Introduction} 
Artificial intelligence (AI) has been extensively applied in the physical layer of wireless communication systems. AI-based technologies have been developed for coding \cite{DEEPTURBO2019}, modulation\cite{OFDM2022}, channel estimation\cite{CE2019}, signal detection\cite{ae2017}, among other applications\cite{AI4PHY2024}. Notably, O’Shea and Hoydis\cite{ae2017} were among the first to introduce an autoencoder (AE) framework (referred as vanilla AE), in which the encoder neural network (NN) functions as the transmitter and the decoder NN as the receiver. This approach conceptualizes communication system design as an end-to-end reconstruction task. Unlike the conventional communication systems that are designed on a block-wise basis with each block optimized individually, the end-to-end AE communication systems use deep learning to jointly optimize the entire system, achieving global optimization.

Subsequently, AE-based systems have been applied to various tasks\cite{sadeghi2019physical}\cite{lu2023attention}\cite{bo2024joint}, such as generating signals with specific properties. For instance, these systems have been used for constellation design \cite{ma2021joint} and for producing OFDM signals with a low peak-to-average power ratio (PAPR) \cite{PAPR2018}\cite{OFDM2022}, all while maintaining robust signal detection performance. Despite the numerous studies on end-to-end AE communication systems, few have noticed their potential in security applications, such as generating noise-like signals that are difficult to be detected by third parties\cite{lopez2005digital}.

This paper aims to design an AE-based communication system that generates featureless signals with low probability of detection and interception (LPD/LPI) \cite{lopez2005digital}, while maintaining Block Error Rate (BLER) performance. A featureless signal is one that resembles noise, with no distinguishable characteristics.

Two related studies by Shakeel \textit{et al.} separately investigate AE-based and traditional block-wise approaches to featureless signal design \cite{shakeel2018machine}\cite{featureless2023}, motivated by the periodic spikes in the autocorrelation functions (ACF) of direct-sequence spread-spectrum (DSSS) signals \cite{wei2022detection}. The first study \cite{shakeel2018machine} employs an AE to generate featureless signals and evaluates their properties through: 1) non-repetitive constellation plots, 2) Gaussian-like distributions, and 3) low ACF values. However, this AE architecture follows \cite{ae2017}, with one fewer layer in the receiver and no explicit mechanism to enforce featurelessness. The author merely claims that training on an additive white Gaussian noise (AWGN) channel yields noise-like signals. Besides, in \cite{shakeel2018machine}, AE transmitter produces a long complex signal of length \(n/2\) for each \(k\)-bit information block but this setting reduces the information-to-signal ratio and decreases throughput. We observe that given \(k\), reducing \(n\) to increase throughput significantly raises the ACF, compromising featurelessness. The second study \cite{featureless2023} uses a block-wise design, without AI, to produce Gaussian-distributed signals for covert communication, but this design suffers from poorer BLER.

To reduce the feature (in terms of ACF) while keeping information-to-signal ratio and throughput high, this study proposes a novel training mechanism for AE-based systems. Specifically, we use Kullback-Leibler (KL) divergence\cite{kingma2013auto} to measure the difference between the distribution of AE-generated signals and a Gaussian distribution and incorporate this KL divergence loss into the conventional categorical cross-entropy (CE) loss. The AE trained with this new loss reduces ACF even for small \(n\) (relative to a fixed small \(k\)), which achieves high throughput and comparable BLER performance. Although KL divergence loss has been applied in variational AE settings \cite{lin2020variational}\cite{VAE_KL2024} to improve training robustness, we are the first to use it in AE to generate noise-like signals.

Another approach to achieving featureless signals while maintaining a high information-to-signal ratio and throughput is to use both a large AE signal length \(n/2\) and a large source block length \(k\). However, in vanilla AE, each \(k\)-bit information block is encoded as a one-hot vector of length \(2^k\), making the network complexity grow exponentially with \(k\). Although \cite{shakeel2018machine} explored directly using the raw \(k\)-bit binary vector as input to reduce complexity, this approach degrades BLER performance compared to the one-hot input setting. 

To improve BLER in AEs with binary input, we introduce a new AE-based architecture that integrates error-correction coding (ECC). The raw \(k\)-bit information block is first ECC-encoded, and the resulting coded bits are used as input to the transmitter NN. The receiver network outputs predicted coded bits, which are then decoded conventionally to recover the information block. We observe that this ECC integration enables the AE to learn structured input patterns, improving coded-block BLER, while final decoding further reduces source-block BLER. Although \cite{codedInput2022} explores pre-coded AE inputs, their method is limited to unity-rate codes with iterative decoding and does not address signal featurelessness. In contrast, our approach supports general ECC schemes and trains the AE with the proposed loss to enhance featurelessness while maintaining high throughput and BLER performance. 

Finally, we deploy the AE-based communication system in a real over-the-air environment and observe that the BLER for information blocks is zero.

The main contributions of this study are: 1) We design a novel loss by integrating a KL divergence term with the categorical cross‐entropy loss, making the AE generated signals more featureless. 2) We substitute one-hot inputs with binary inputs pre-encoded by the error-correction coding scheme and apply this scheme to decode the AE’s outputs, improving BLER on both coded and decoded blocks. 3) We demonstrate the viability of our AE-based system in the real-world wireless communication. 

The paper is structured as follows. Section~\ref{lbl:basedline} analyses the baseline AE systems and highlights their drawbacks. Section~\ref{lbl:proposed} proposes our AE system. Section~\ref{lbl:experiment} shows simulation results. The paper is concluded in Section~\ref{lbl:conclusion}.

\section{Baseline Autoencoder-based solutions}\label{lbl:basedline}


\subsection{The vanilla Autoencoder} 
\begin{figure}[H]
    \centering
    \includegraphics[width=0.85\linewidth]{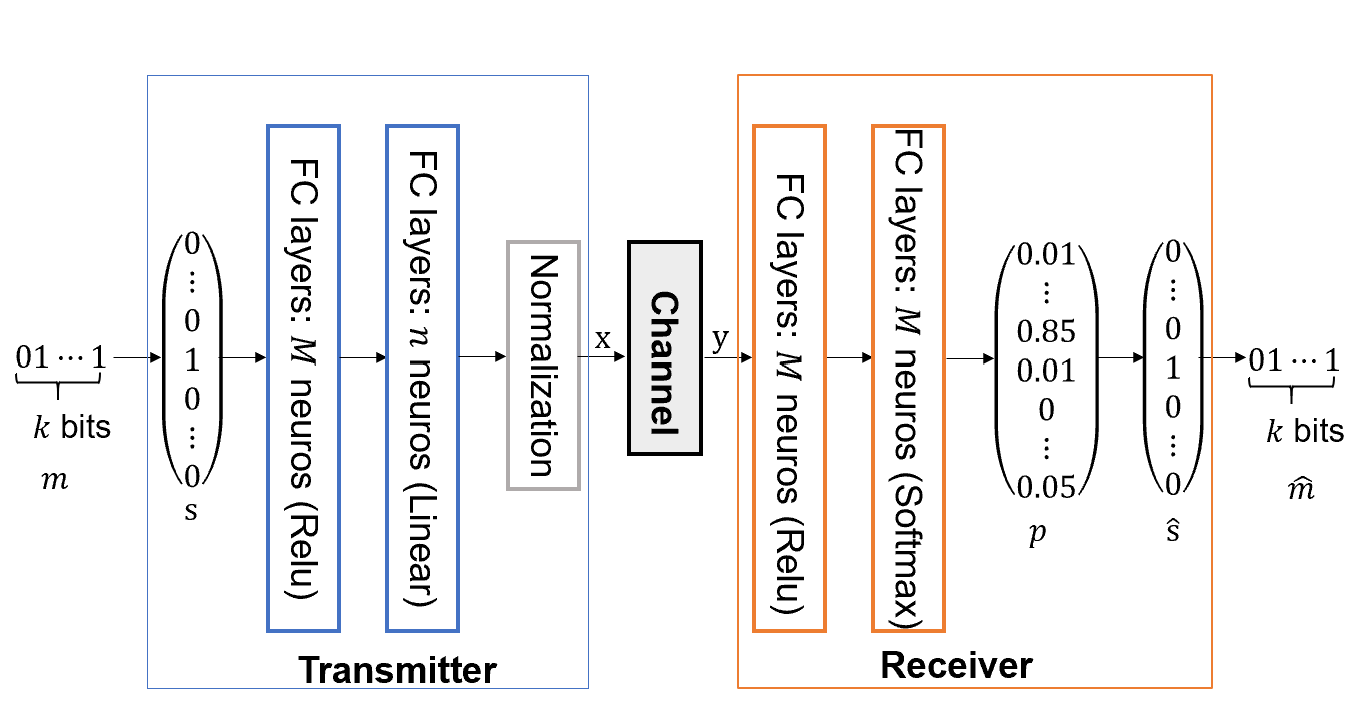}
    \caption{The vanilla AE communication system.}
    \label{AE struc1}
\end{figure}

The basic architecture of vanilla AE\cite{ae2017} is shown in Figure \ref{AE struc1}. It utilizes a feedforward neural network-based transmitter and receiver. The transmitter consists of two fully connected (FC) layers, with the input of a one-hot vector $ s $with length of  $M=2^k$ generated from a $ k $-bit message BLOCK. The second layer in the transmitter NN produces $ n $real-value outputs, which is then converted to the complex-value signal  $ \bold x $of length $ n/2 $. The signal $ \bold x $is normalized to satisfy the average power constraint $\mathbb{E} \left[ |x_i|^2 \right]\leq1,\forall i$.  After receiving $ \bold y $from the channel, the receiver consists of two dense layers, with a softmax activation in its final layer, producing a probability vector $ p $in $(0,1)^M$  over all possible messages. The decoded message $ \hat{s} $is determined by selecting the index of the highest probability element in $ p $. Then the predicted one-hot vector $ \hat{s} $

The basic architecture of vanilla AE\cite{ae2017} is shown in Figure \ref{AE struc1}. It utilizes a feedforward neural network-based transmitter and receiver. The transmitter consists of two fully connected (FC) layers, with the input of a one-hot vector $ s $with length of  $ M=2^k $generated from a $ k $-bit message BLOCK. The second layer in the trasmitter NN produces $ n $real-value outputs, which is then converted to the complex-value signal  $ \bold x $of length $ n/2 $. The signal $ \bold x $is normalized to satisfy the average power constraint $\mathbb{E} \left[ |x_i|^2 \right]\leq1,\forall i$.  After receiving $ \bold y $from the channel, the receiver consists of two dense layers, with a softmax activation in its final layer, producing a probability vector $ p $in $(0,1)^M $over all possible messages. The decoded message $ \hat{s} $is determined by selecting the index of the highest probability element in $ p $. Then the predicted one-hot vector $ \hat{s} $
is converted to predicted binary message $ \hat{m} $with $ k $-bit. Meanwhile, for the AE used in \cite{shakeel2018machine}, there are two dense layers for transmitter and  only one dense layers for receiver.

In the vanilla AE, the softmax is used for the output layer activation. Given the input vector $ z $with length $ M$, each element $ p_i $($ i=1,2,...M$) of the output $ p $of the softmax function is calculated as

\begin{equation}
p_i = \frac{e^{z_i}}{\sum_{j=1}^{M} e^{z_j}}, \quad \text{for } i = 1, \dots, M
\end{equation}
To minimize the detection error rate, the loss function employs the categorical cross entropy, which is defined as 
\begin{equation}
L_{\text{CE}}  = - \sum_{i=1}^{M} s_i \log(p_i)
\end{equation}
The loss is minimized by the  stochastic
gradient descent (SGD) algorithm during the end-to-end training of the AE.

\subsubsection{performance investigation} We present the ACF values and BLER performance of the vanilla AE with various value of \( n \). The simulation settings are as follows: We randomly generate 1,000 message blocks, each consisting of \( k = 4 \) bits. These blocks are converted into one-hot vectors and used as input to the AE. The channel is modeled as an AWGN channel. Training is performed at a fixed \( E_b/N_0 \) value of 0 dB using the Adam optimizer with a learning rate of 0.001. 

\begin{figure}[H]
    \centering
    \begin{subfigure}[b]{0.48\linewidth}
        \centering
        \includegraphics[width=\linewidth]{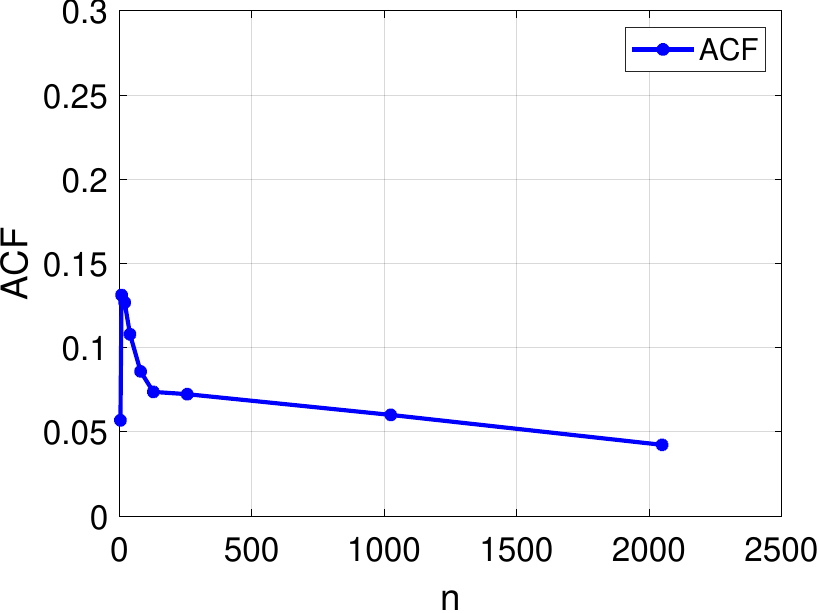}
        \caption{\vspace{2em} ACF}
        \label{fig1-ACFAE}
    \end{subfigure}
    \hfill
    \begin{subfigure}[b]{0.47\linewidth}
        \centering
        \includegraphics[width=\linewidth]
        {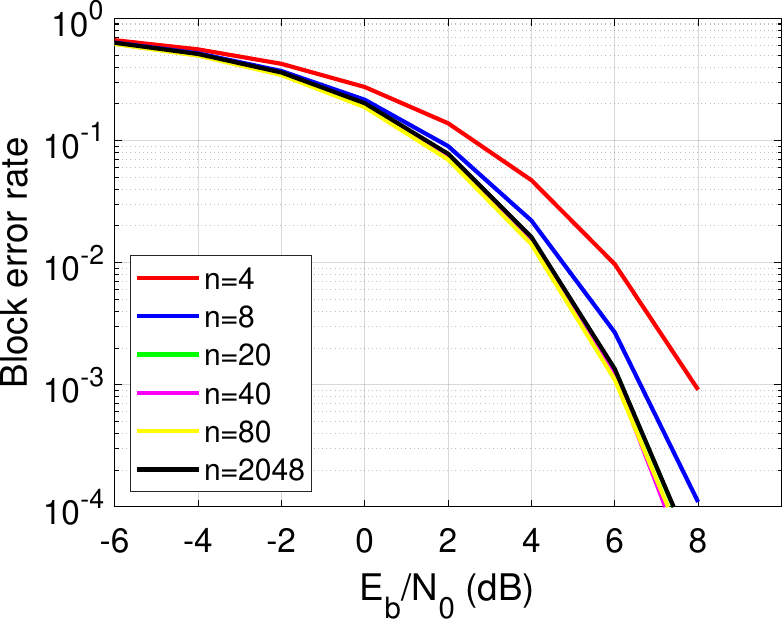}
        \caption{\vspace{2em} BLER}
        \label{fig:AE perf1}
    \end{subfigure}
    \caption{Performance of vanilla AE on AWGN channel: \(k=4\)}
    \label{fig:AE perf}
\end{figure}

We firstly analyze the impact of \( n \) on the performance, considering \( n \in \{4, 8, 20, 40, 80, 128, 256, 1024, 2048\} \). Shakeel’s work \cite{shakeel2018machine} only uses large \( n \), which is 256 or 2048, to generate signals with very low ACF. However, higher \( n \) with lower information-to-signal ratio causes lower throughput. Figure \ref{fig:AE perf}-(a) presents the ACF values for these values of \( n \). Given an AE signal transmitted \( \boldsymbol{x} \) with 1,000 symbols, the ACF is computed as the highest absolute ACF value among different values of lag, ranging from 1 to 50. The point is plotted by taking average on the highest absolute ACF values of 10,000 AE signals. It is observed that low ACF values occur when \( n \) is higher than 1024 or equal to 4. However, for \( n = 8,   20,  \) or \( 40 \), the ACF is very high, exceeding 0.1.  

Meanwhile, Figure \ref{fig:AE perf}-(b) shows that the BLER is high when \( n = 4 \). As \( n \) increases from 4 to 20, the BLER decreases, but for \( n > 20 \), the reduction in BLER becomes minimal. Due to the space of the figure, the BLER results of AE with \( n = 128, 256, \) and \( 1024 \) are not shown in the figure. These BLER values are also very similar to those of the AE with \( n = 80 \) and \( 2048 \).  

Taking into account both low ACF and low BLER, the optimal choice for \( n \) is 2048 among these investigated values. However, this high value of \( n \) significantly reduces throughput. Considering both BLER and throughput, \( n =20\) is preferable, but this leads to an undesirably high ACF, which makes the signal not like the noise signal. 

Therefore, the first important problem occurs: how to effectively make the AE generated signal more featureless while maintaining low BLER and high throughput?

\subsection{Autoencoder with direct \(k\)-bit Input}\label{sec:kdirectbit}
Figure~\ref{AE struc1} shows that a information block of size \(k\) bits requires \(2^k\) input neurons, which quickly becomes impractical as \(k\) grows. To address the infeasibility of the vanilla AE for large information blocks, two strategies can be employed. First, one can split the block into several smaller sub-blocks—at the cost of reduced throughput. Second, as suggested in \cite{shakeel2018machine}, the raw \(k\)-bit binary vector directly serves directly as the autoencoder’s input. Although this latter approach is feasible for larger \(k\), it shifts from predicting a single message index to predicting each bit individually, thereby increasing the complexity of signal detection.
\subsubsection{performance investigation} 
\begin{figure}[H]
    \centering
    \begin{subfigure}[b]{0.48\linewidth}
        \centering
        \includegraphics[width=\linewidth]{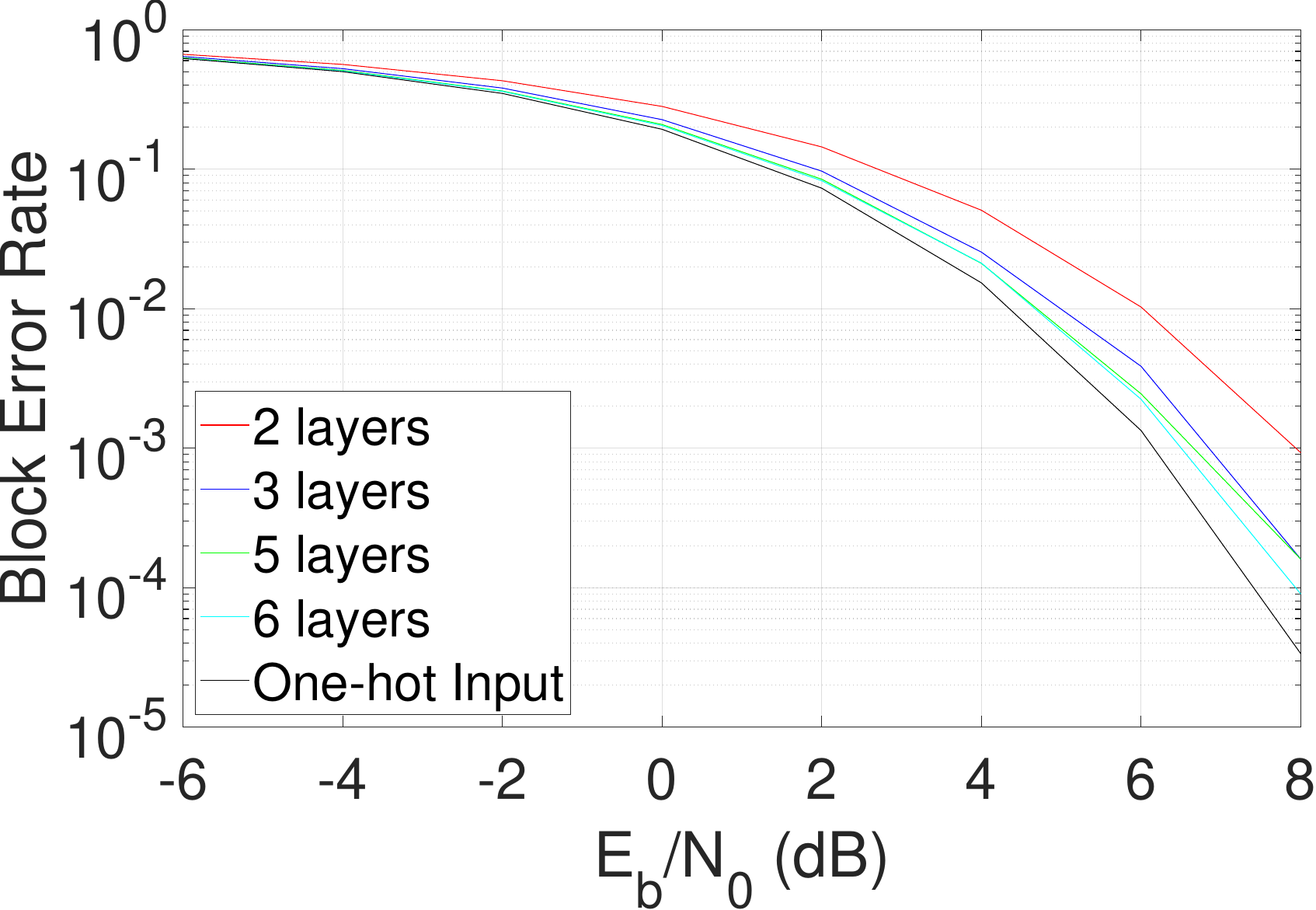}
        \caption{\vspace{2em} k-bit input}
        \label{fig1-ACFAE}
    \end{subfigure}
    \hfill
    \begin{subfigure}[b]{0.46\linewidth}
        \centering
        \includegraphics[width=\linewidth]
        {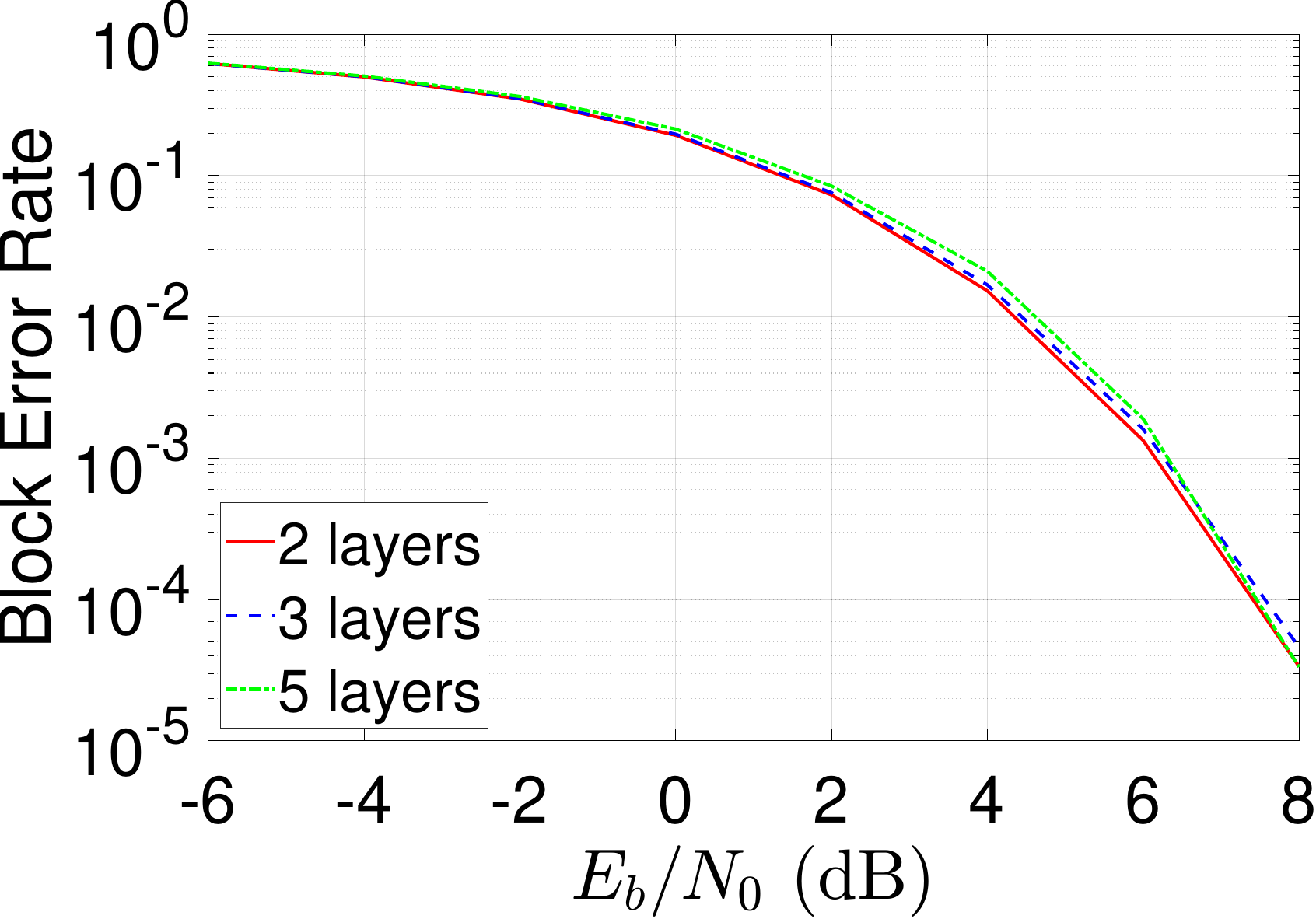}
        \caption{\vspace{2em} One-hot input}
        \label{fig:AE perf1}
    \end{subfigure}
    \caption{Performance of AE with different layers}
    \label{fig5-layers}
\end{figure}

We assess the BLER of an AE(\( k=4,n=20\)) that takes a direct $k$-bit binary input instead of a one-hot input of length $M = 2^k$. An “$N$-layer AE” denotes $N$ layers in both the transmitter and receiver. As shown in Figure~\ref{fig5-layers}-(a), the 2-layer AE with $k$-bit binary input performs much worse than the vanilla AE.

To address this, we add hidden layers—each of width $M$—to both transmitter and receiver NN. Figure~\ref{fig5-layers}-(a) shows BLER gains when increasing from 2 to 3 layers, with minor improvements up to 5 layers and negligible benefit beyond. Nevertheless, even a 5-layer binary-input AE underperforms the vanilla AE with one-hot input. Figure~\ref{fig5-layers}-(b) shows that increasing layers for one-hot input does not improve the BLER performance. This raises the second key problem: How can higher BLER performance be achieved with a thin AE using direct $k$-bit inputs?

\section{Our proposed Autoencoder MODEL}\label{lbl:proposed}

\subsection{Proposed method to enhance featureless property}
To address the first problem that how to make the AE signal more featureless while maintaining low BLER and high throughput, we propose a new method to train AE in this part.

It is known that KL divergence \cite{kldiv} is a statistical distance that quantifies the difference between a model's probability distribution and the true probability distribution. It has been widely applied in various fields, such as coding theory, statistics, and inference. Here, we utilize the KL divergence to measure the discrepancy between the distribution of AE generated signals and a Gaussian distribution. We propose a new loss to train the AE by incorporating this KL divergence into the previous CE loss.

Specifically, we first calculate the KL divergence between the distribution of \(q(x|m\) and a prior \(p(a)\), which follows the Gaussian distribution of \(p(a) \sim \mathcal{N}(0, I)\). The approximated KL divergence formula is  defined as 
\begin{equation}
L_{\text{KLD}} \left( q(x|m), p(a) \right) = \frac{1}{2}  \left( 1 + \log \sigma^2 - \mu^2 - \sigma^2 \right)
\end{equation}
where \(\mu\) and \(\sigma^2\) are separately the mean and variance of transmitted signal \( \bold x \). Detailed derivations are shown in \cite{kingma2013auto}. Then, we add this KL loss to the CE loss. Therefore, the new loss function of the AE is 
\begin{equation}
L = L_{\text{CE}} + \alpha*L_{\text{KLD}}
\end{equation}
where \(\alpha\) is a hyperparameter. This proposed loss provides an explicit mechanism to make the signal more noise-like while keeping the BLER as low as possible.

\subsection{The Proposed AE-based system with binary input} 
\begin{figure}[H]
    \centering
    \includegraphics[width=0.9\linewidth]{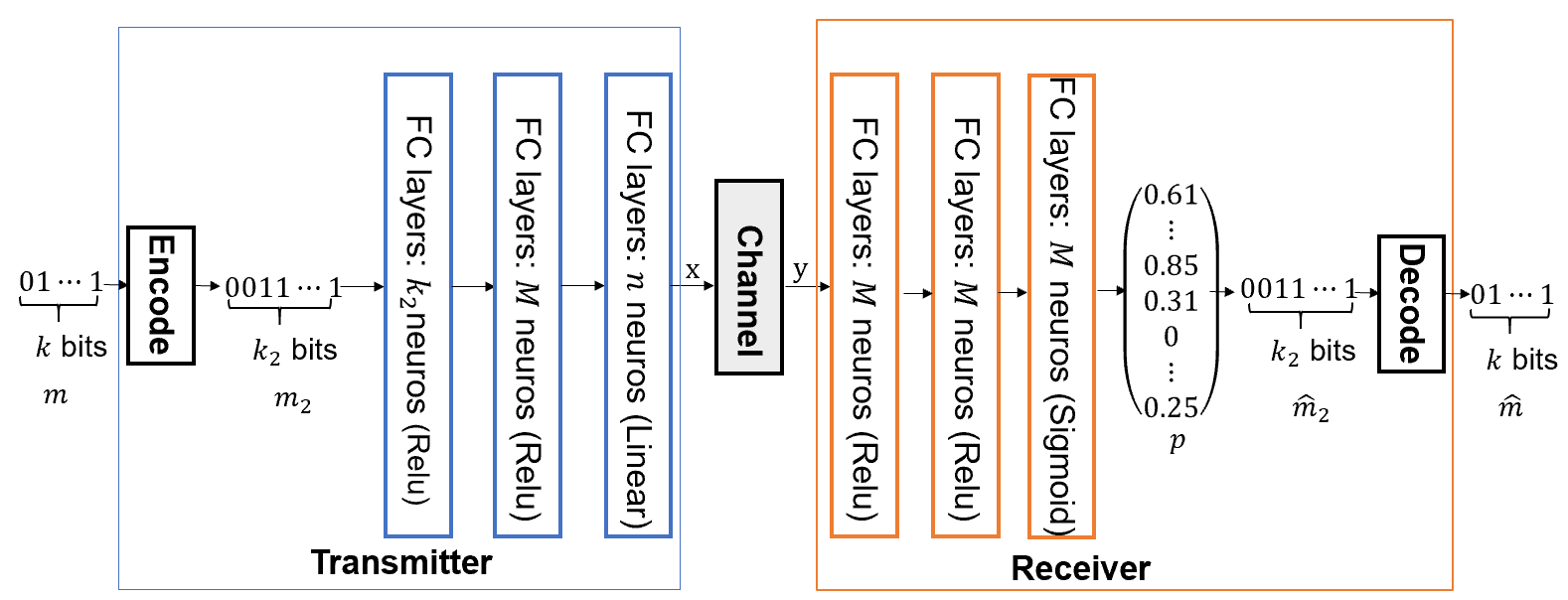}
    \caption{The proposed AE communication system with coded input.}
    \label{AE struc-codebi}
\end{figure}

To  solve the second problem that AE with direct \( k \)-bit input has lower BLER performance than the vanilla AE, we propose the solution illustrated in Figure \ref{AE struc-codebi}. Specifically, the \( k \)-bit input \( m \) is encoded into \( m_2 \) with \( k_2 \) bits using ECC, which is then used as the input of the AE. Since the task is more complex than vanillar AE, we use 3 layers in both transmitter and receiver networks. 

Once encoded, the binary input \( m_2 \) is passed through FC layers in the transmitter part, generating \( n \) real values that represent a complex-valued  \( \bold x \) of length \( n/2 \). Upon receiving \( y \) from the channel, the receiver—comprising several FC layers—attempts to reconstruct the pre-coded message bits. Unlike the vanilla AE, which uses a softmax activation function, we employ a sigmoid activation function to predict each bit within the block. The predicted \( k_2 \)-bit coded block \( \hat{m_2} \) is then decoded using the conventional coding mechanism to obtain the final predicted \( k \)-bit information block \( \hat{m} \). Since the chosen coding mechanism has error correction capabilities, it further enhances the BLER performance of the raw \( k \)-bit block. Besides, to simultaneously enhance the featureless property and keep the BLER performance, this proposed AE is trained with the proposed loss.

\section{Experimental results}\label{lbl:experiment}

\subsection{Performance of the proposed method to enhance featureless property}
\subsubsection{ACF and BLER analysis}
In this subsection, we evaluate the ACF and BLER performance of the AE trained with the proposed loss function. The structure of vanilla AE with one-hot input is employed. Both AWGN channel and Rayleigh fading channel are tested. 
\begin{figure}[H]
    \centering
    \begin{subfigure}[b]{0.48\linewidth}
        \centering
        \includegraphics[width=\linewidth]{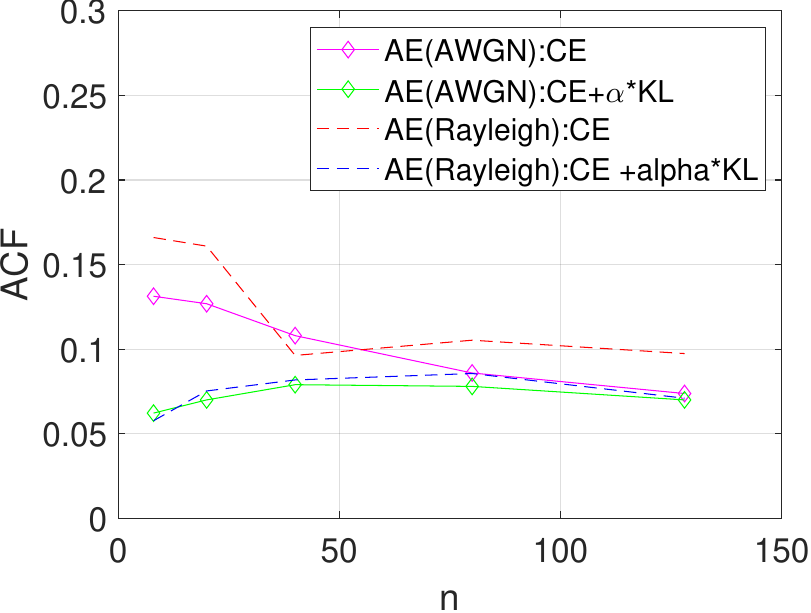}
        \caption{\vspace{2em} ACF of proposed method.}
    \end{subfigure}
    \hfill    
    \begin{subfigure}[b]{0.48\linewidth}
        \centering
        \includegraphics[width=\linewidth]{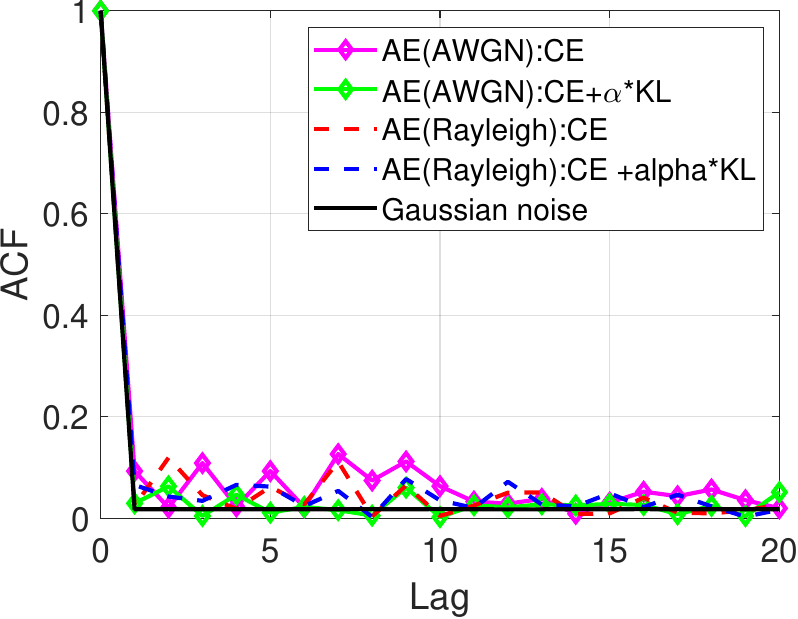}
        \caption{\vspace{2em} ACF on each Lag}
        \label{fig3-1-ACF_AE_KL}
    \end{subfigure}
    \caption{ACF of AE (\(k=4\))}
    \label{fig4-ACF_AE_KL}
\end{figure}

\begin{figure}[H]
    \centering
    \begin{subfigure}[b]{0.48\linewidth}
        \centering
        \includegraphics[width=\linewidth]
        {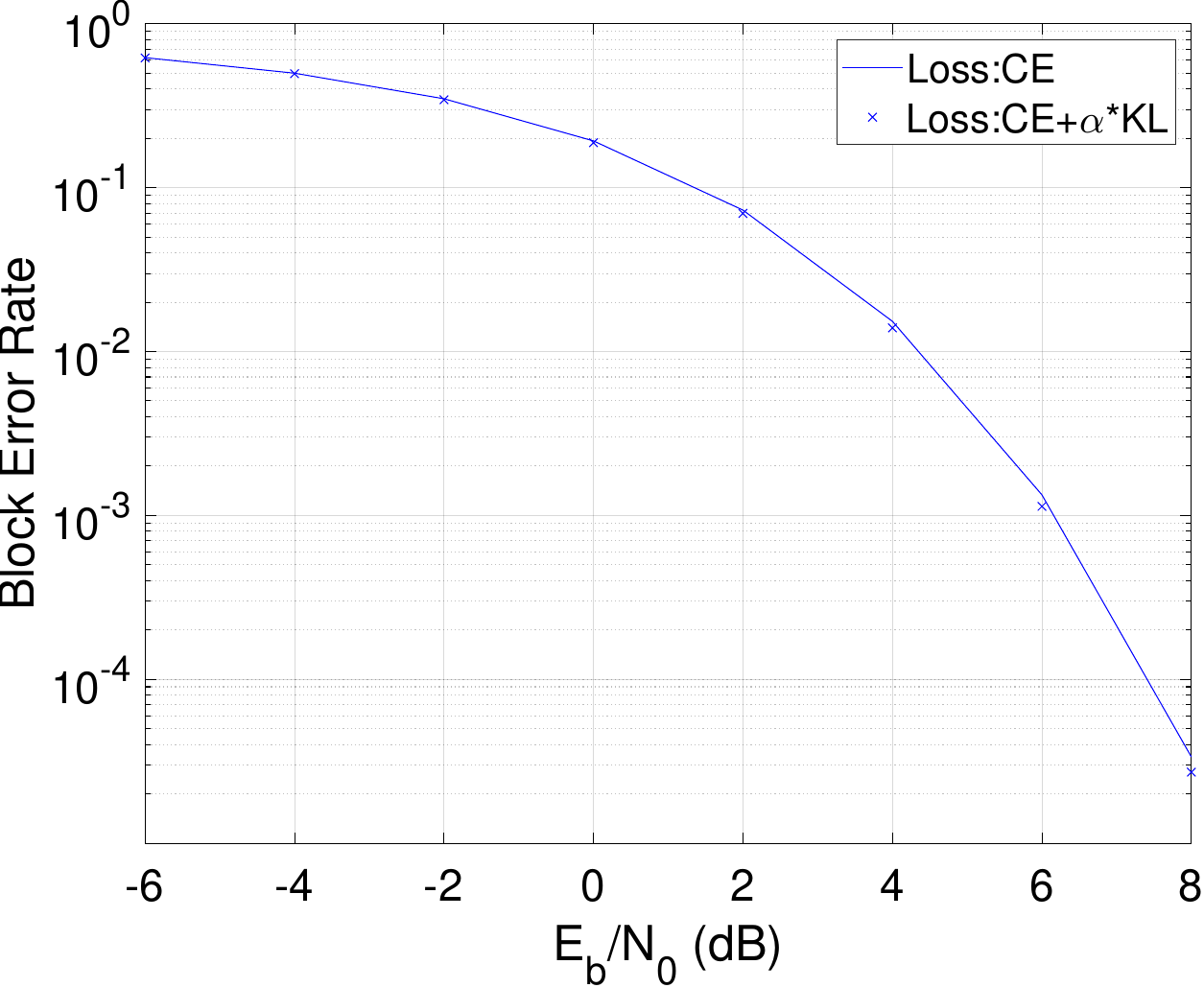}
        \caption{\vspace{2em} AWGN channel}
        \label{fig3-2BLER}
    \end{subfigure}
    \hfill
    \begin{subfigure}[b]{0.48\linewidth}
        \centering
        \includegraphics[width=\linewidth]
        {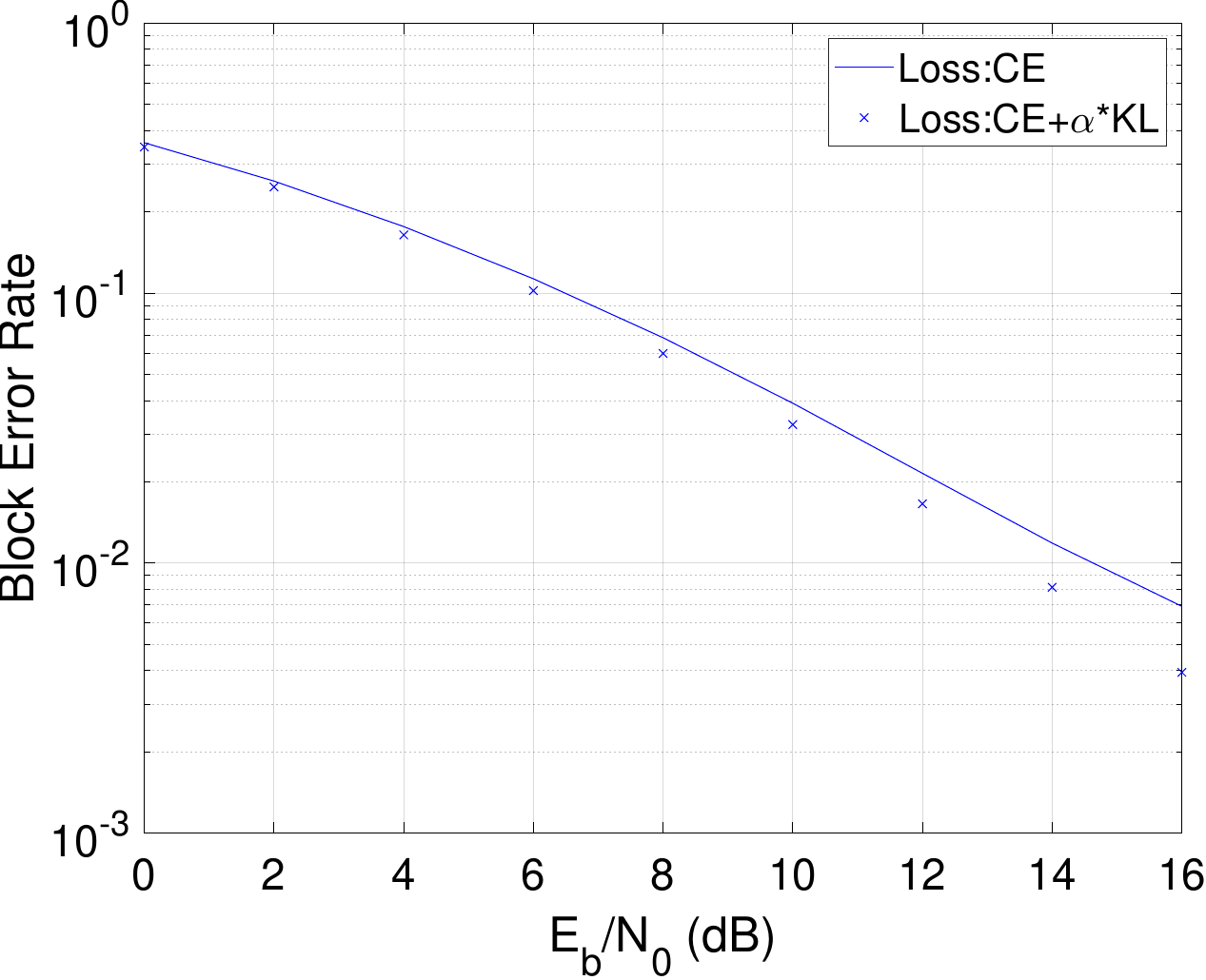}
        \caption{\vspace{2em} Rayleigh fading channel}
        \label{fig3-2BLER}
    \end{subfigure}
    \caption{BLER of AE (\(k=4, n=20\)) with proposed loss}
     \label{fig3-1-ACF_AE_KL}
\end{figure}

 We first present the ACF performance. Figure \ref{fig4-ACF_AE_KL}-(a) shows that, compared to the previous loss function, the ACF is reduced for each value investigated of \( n \in \{8, 20, 40, 80, 128\} \). In particular, for \( n = 8, 20,\) or \(40\), the ACF is significantly lower than before. While considering low ACF, low BLER and high throughput, \( n =  20\) is most suitable setting among these investigated values for \( k = 4\). Moreover, for the AE with \( k = 4 \) and \( n = 20 \), Figure \ref{fig4-ACF_AE_KL}-(b) shows that the AE trained with the proposed loss function achieves lower ACF values for most lag values ranging from 1 to 20. Moreover,
 Figure \ref{fig3-1-ACF_AE_KL} shows that the BLER performance of AE with \( k = 4 \) and \( n = 20 \) remains similar between the two loss functions. This suggests that the proposed loss effectively reduces the ACF, making the signal more noise-like, while maintaining the BLER performance.
\subsubsection{Featureless property analysis}
As mentioned earlier, Shakeel’s work explores the featureless properties of AE-generated signals with large-structure AE \((k=32,n=2048)\) with one-hot input through three key aspects: 1) the constellation plot exhibits no repetitive patterns; 2) the signal resembles Gaussian noise in its distribution; and 3) the ACF of the signal is small. 
\begin{figure}[H]
    \centering
    \begin{subfigure}[b]{0.41\linewidth}
        \centering
        \includegraphics[width=\linewidth]{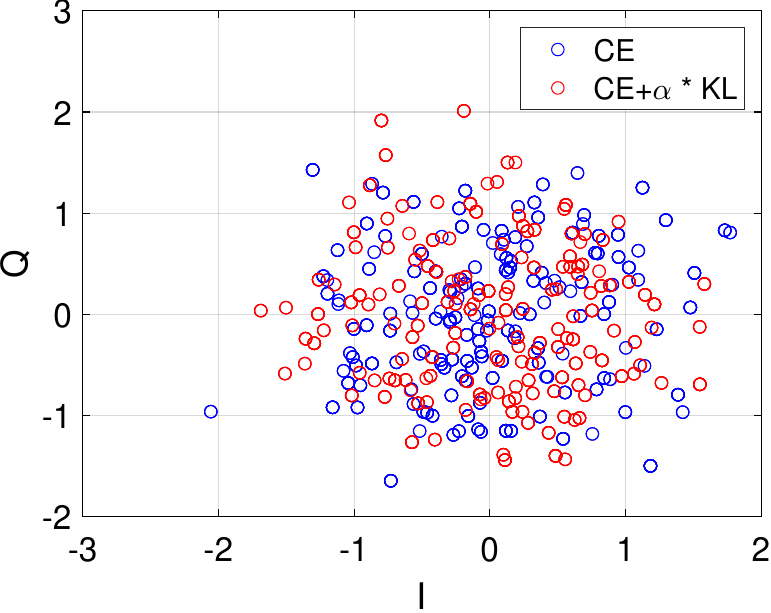}
        \caption{\vspace{2em} Constellation}
        \label{fig91Constellation}
    \end{subfigure}
    \hfill
    \begin{subfigure}[b]{0.43\linewidth}
        \centering
        \includegraphics[width=\linewidth]{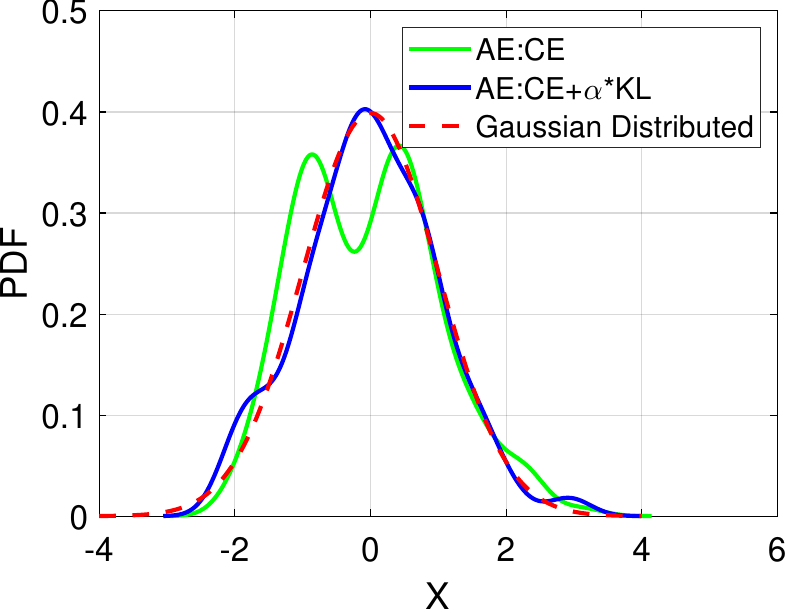}
        \caption{\vspace{2em} Distribution}
        \label{fig92ACF}
    \end{subfigure}
    \hfill
    \begin{subfigure}[b]{0.7\linewidth}
        \centering
        \includegraphics[width=\linewidth]{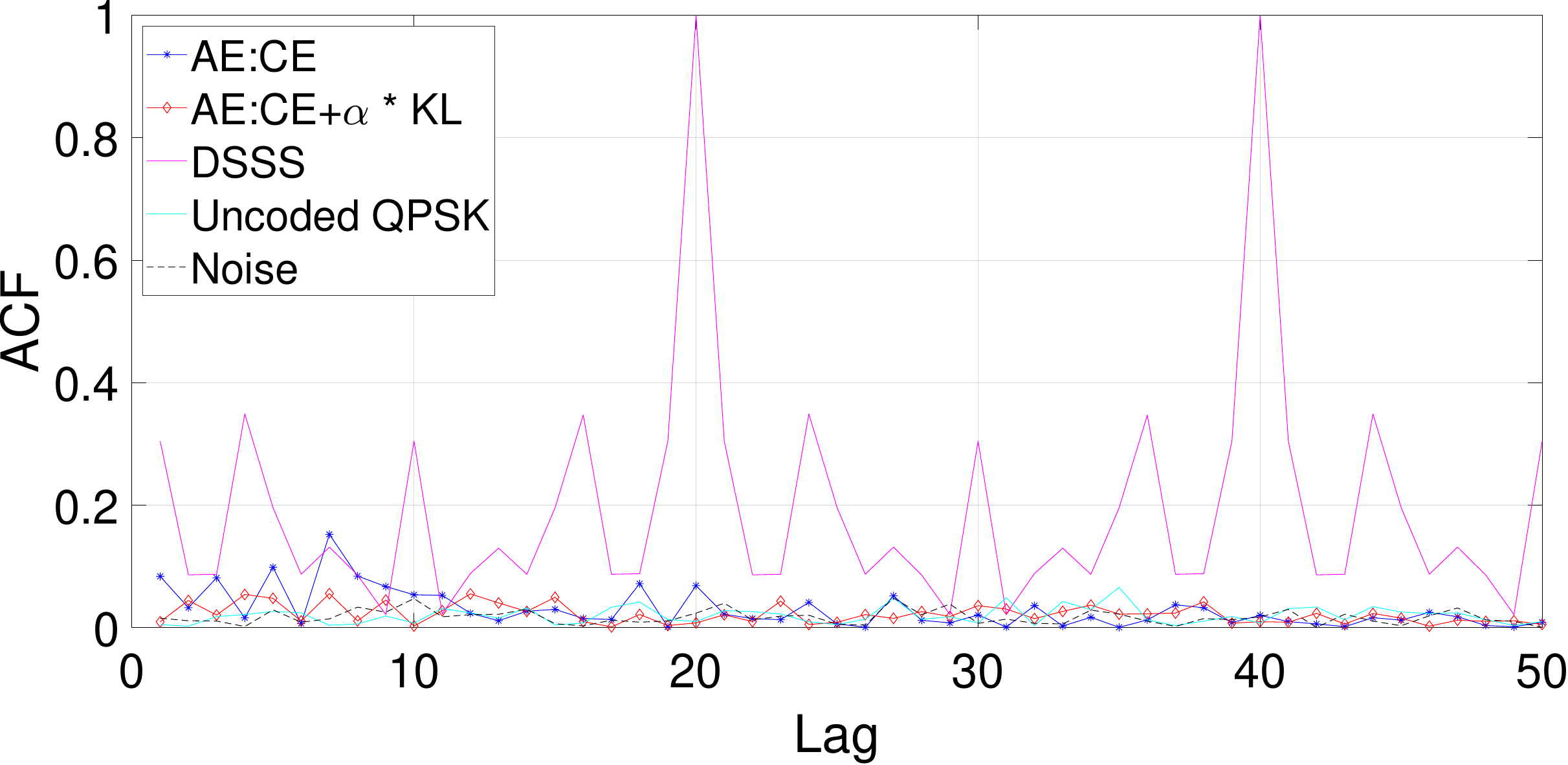}
        \caption{\vspace{2em} ACF}
        \label{fig92ACF}
    \end{subfigure}
    \caption{Featureless properties of AE (\(k=4, n=20\)) signal}
    \label{Featureless property}
\end{figure}

Next, we investigate these three properties of the AE ($k=4,n=20$) trained with our proposed loss function. The results, obtained over an AWGN channel (with similar propertiess under Rayleigh fading), are as follows. First, Figure~\ref{Featureless property}-(a) compares the constellations of AEs trained with the CE-only loss and with the proposed loss, showing that both AEs produces an irregular, featureless constellation. Figure~\ref{Featureless property}-(b) demonstrates that the AE with proposed loss generates symbols whose distribution closely matches a standard Gaussian, whereas the AE signal with CE-only loss exhibits a bimodal distribution. Finally, Figure~\ref{Featureless property}-(c) presents the ACFs among Lags ranging from 1 to 50. The DSSS signal shows periodic peaks, while the AE with proposed loss yields much smaller ACF values than AE with CE-only loss. Although the uncoded QPSK with random bits also exhibits low ACF due to symbol independence, its regular constellation pattern is highly distinguishable.

\subsection{Performance of the proposed AE-based system}
\begin{figure}[H]
    \centering
    \begin{subfigure}[b]{0.48\linewidth}
        \centering
         \includegraphics[width=\linewidth]{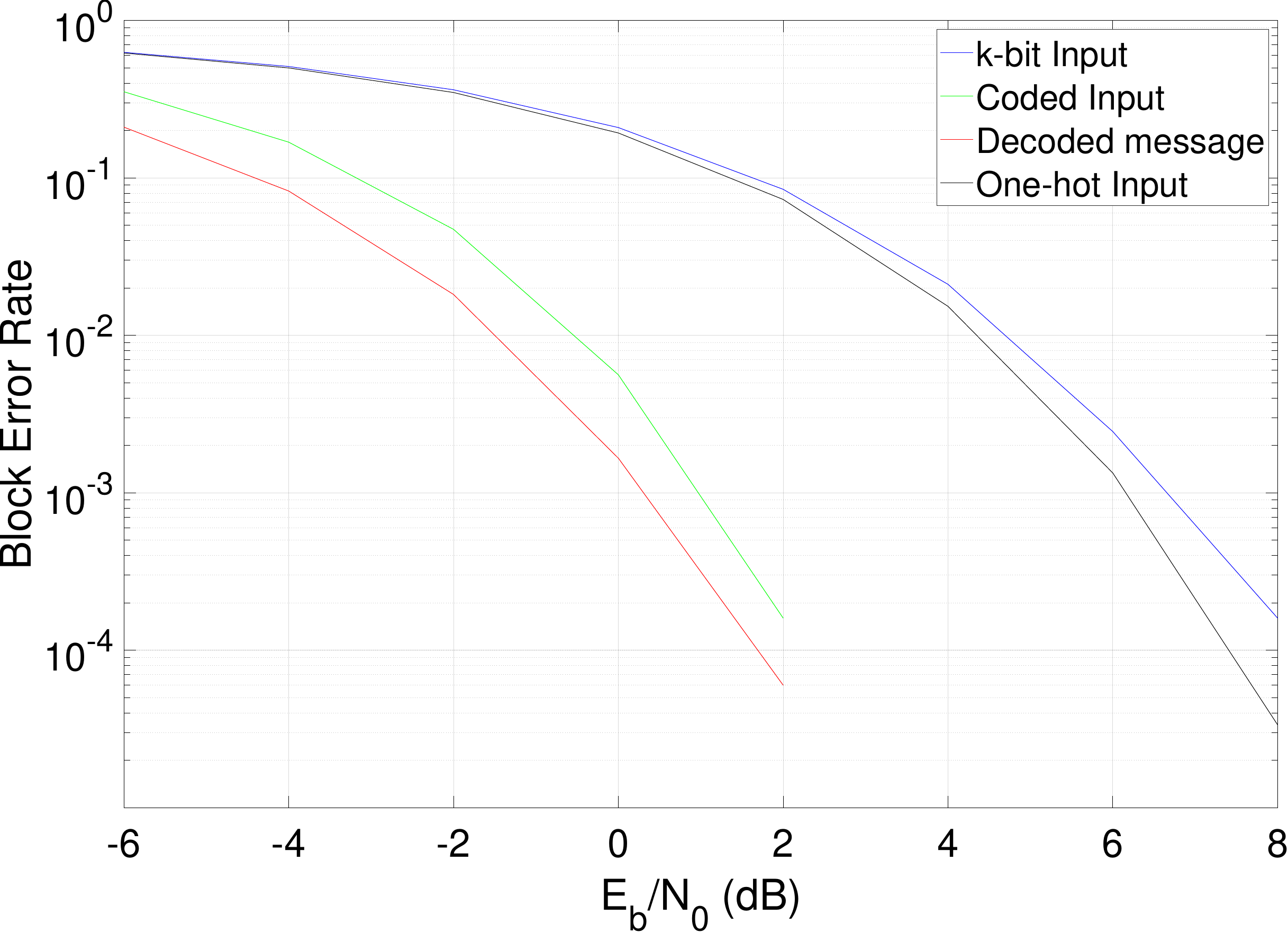}        
        \caption{\vspace{2em} AWGN channel}
           
    \end{subfigure}
    \hfill
    \begin{subfigure}[b]{0.48\linewidth}
        \centering
       \includegraphics[width=\linewidth]{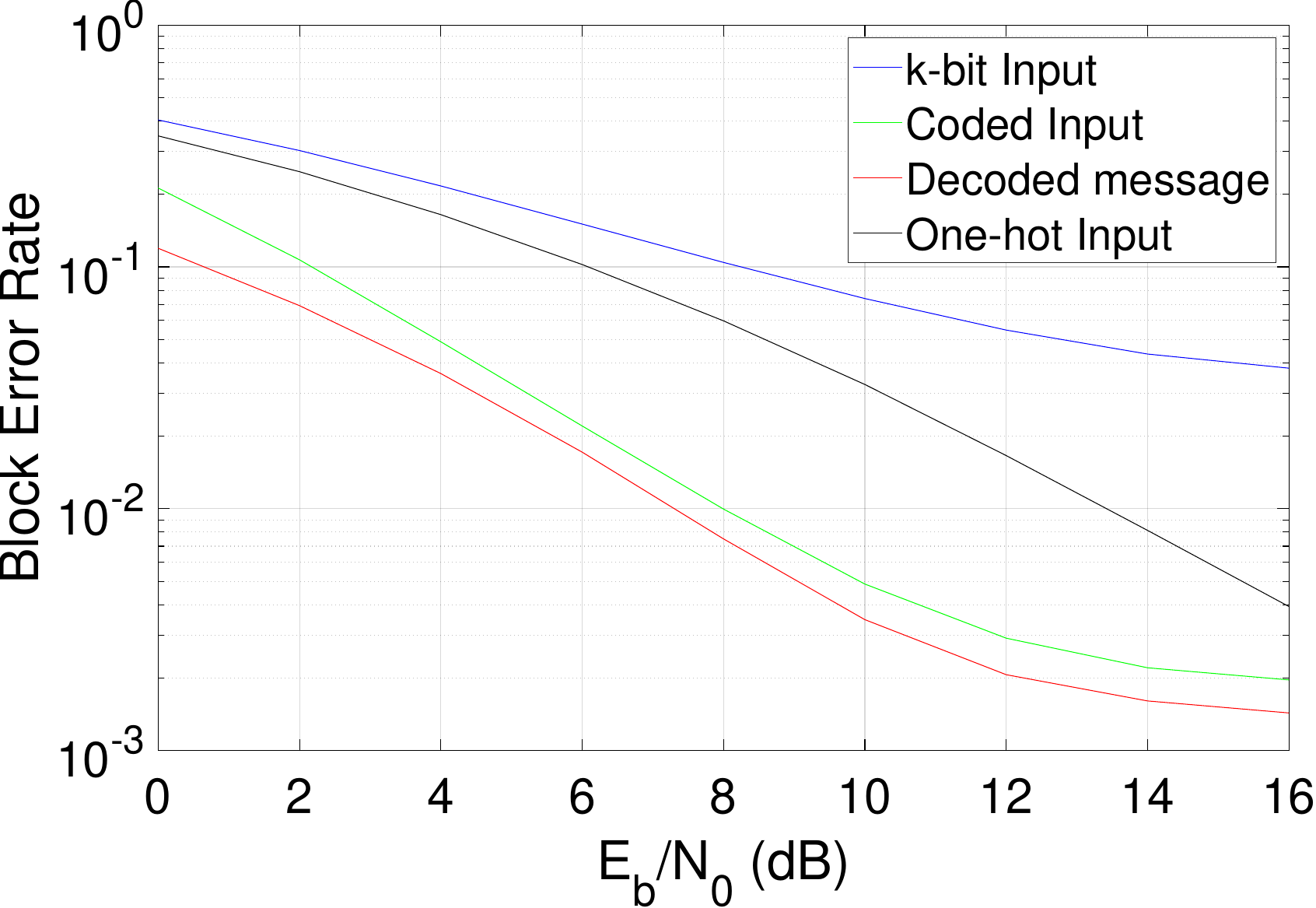}
        \caption{\vspace{2em} Rayleigh fading channel}
      
    \end{subfigure}
    \caption{Performance of AE (\(k_2 = 14, n=20\)) with the BCH (\(k = 4, k_2 = 14\)) coded binary input}
    \label{Fig8-BCH_AE}
\end{figure}

In this section, we analyze the BLER performance of our proposed AE-based system with binary input. Firstly, we implement the Bose–Chaudhuri–Hocquenghem (BCH) coding with a raw \( k \)-bit information block (\( k = 4 \)), resulting a coded input of \( k_2 = 14 \) bits for AE (\(k_2=14, n=20\)). Figure \ref{Fig8-BCH_AE} presents a BLER comparison of 1e5 tested blocks and each block has raw input \( k \) bits. There are four measurement types compared. “\(k\)-bit Input” represents the BLER between \(m\) and \( \hat{m} \) when applying the AE with a direct \(k\)-bit input. “Coded Input” and “Decoded Message” both involve using BCH-coded binary input, where the original \(k\)=4 bits is converted to \(k_2\)=14 bits before being processed by our proposed AE (Figure \ref{AE struc-codebi}). “Coded Input” refers to the BLER between \(m_2\) and \( \hat{m_2} \) of coded blocks while the “Decoded Message” refers to the BLER between \(m\) and \( \hat{m} \) of source message blocks after decoding. “One-hot Input” represents the BLER between \(m\) and \( \hat{m} \) for the vanilla AE (Figure \ref{AE struc1}). 

By comparing “\(k\)-bit Input", “Decoded Message", and “One-hot Input", we can observe that the BLER performance between \( m \) and \( \hat{m} \) for proposed AE with “Decoded Message" is significantly improved on both AWGN channel and Rayleigh fading channel.
Notably, when comparing the “Coded Input" with the “\(k\)-bit Input", and “One-hot Input", we observe an interesting finding: the “Coded Input" has a much lower BLER. This suggests that the coded input provides essential features that enable the AE to effectively learn the coding characteristics and accurately reconstruct the coded binary message in the receiver. Moreover, the “Decoded Message" has better performance compared to the “Coded Input", which is attributed to the error correction capabilities of ECC. 


Furthermore, to evaluate our observations that AE can learn some characteristics of coding schemes, we further investigate the Reed-Solomon (RS) coding and convolutional (Conv) coding for our proposed AE system. As illustrated in Figure \ref{Fig8 RS(9,21,n40)_Conv(10,20,n80)}, AE with coded input using both coding mechanisms outperform the AE of the direct \(k\)-bit input in most scenarios. Besides, the BLER of source message blocks from our proposed AE systems is much lower than this from the AE of the direct \(k\)-bit input. Collectively, these results underscore the effectiveness of the proposed AE communication system in significantly improving the BLER performance.

Moreover, as shown in Figure \ref{Fig7BCH_AE}, the AE in our proposed system trained with the proposed loss attains lower ACF values while maintaining BLER comparable to AE trained with the CE-only loss. These results demonstrate the robustness and suitability of the new loss for binary-input AEs.


\begin{figure}[H]
    \centering
    \begin{subfigure}[b]{0.49\linewidth}
        \centering
         \includegraphics[width=\linewidth]{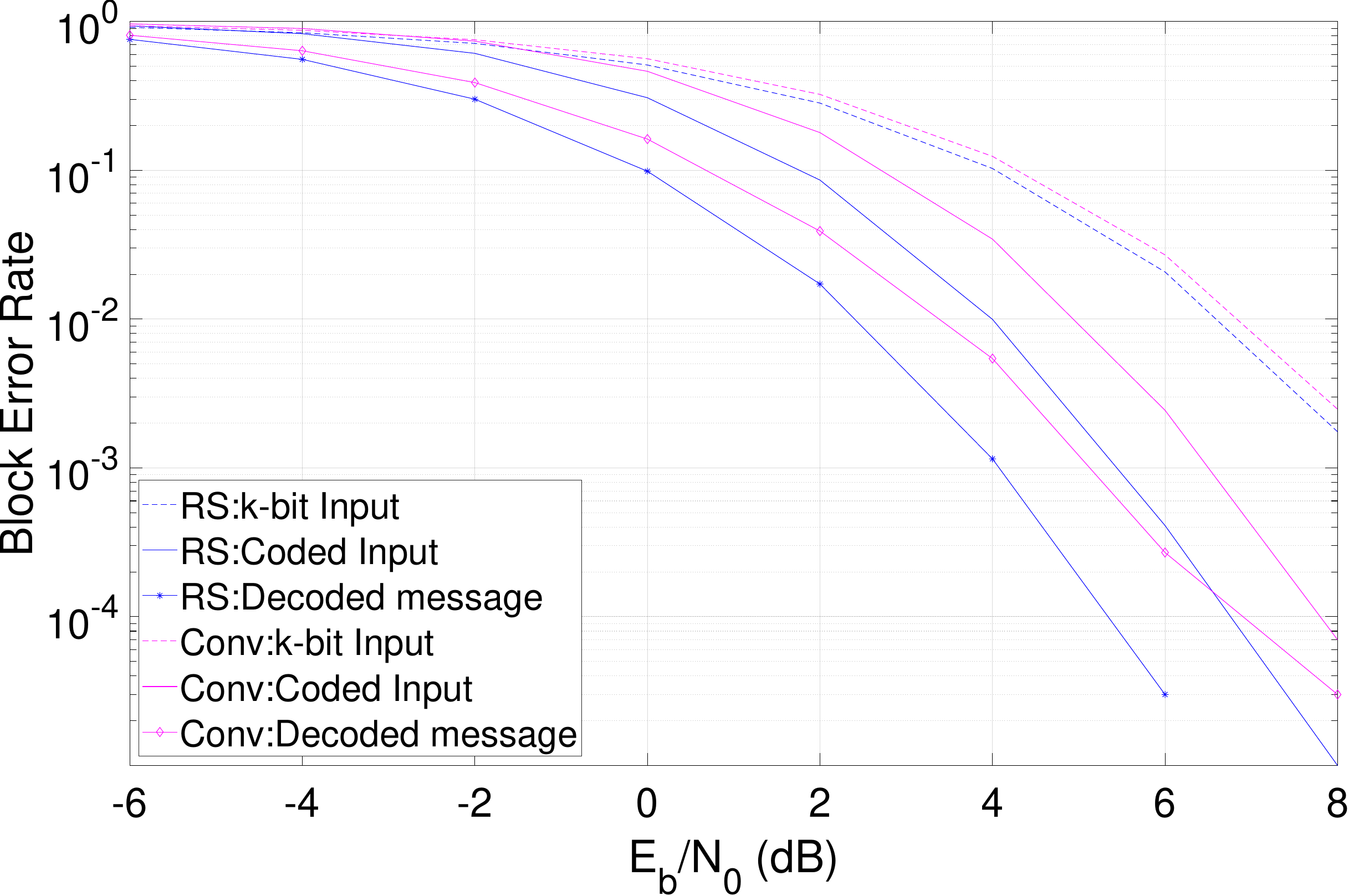}        
        \caption{\vspace{2em} AWGN channel}
        \label{Fig13 sourceImage}       
    \end{subfigure}
    \hfill
    \begin{subfigure}[b]{0.47\linewidth}
        \centering
       \includegraphics[width=\linewidth]{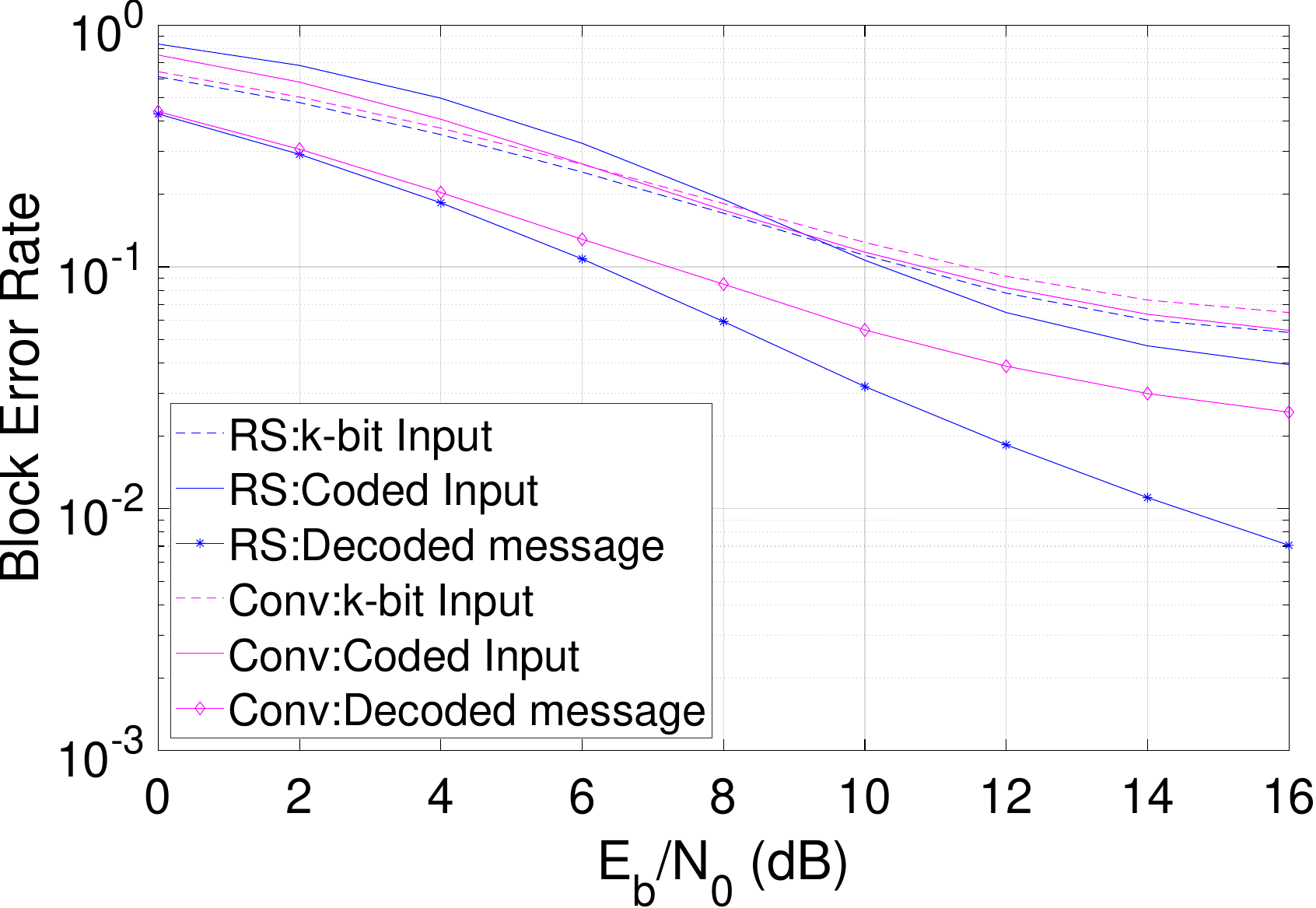}
        \caption{\vspace{2em} Rayleigh fading channel}
        \label{Fig13 RxImage}
    \end{subfigure}
    \caption{Performance of AE with RS (\(k=9\),\(k_2=21\)) coding or Conv (\(k=10\), \(k_2=20\)) coding: For RS coded coding, \(n=40\) is used in AE. For Conv coding, \(n=80\) is used.}
    \label{Fig8 RS(9,21,n40)_Conv(10,20,n80)}
\end{figure}

\begin{figure}[H]
    \centering
    \begin{subfigure}[b]{0.48\linewidth}
        \centering
         \includegraphics[width=\linewidth]{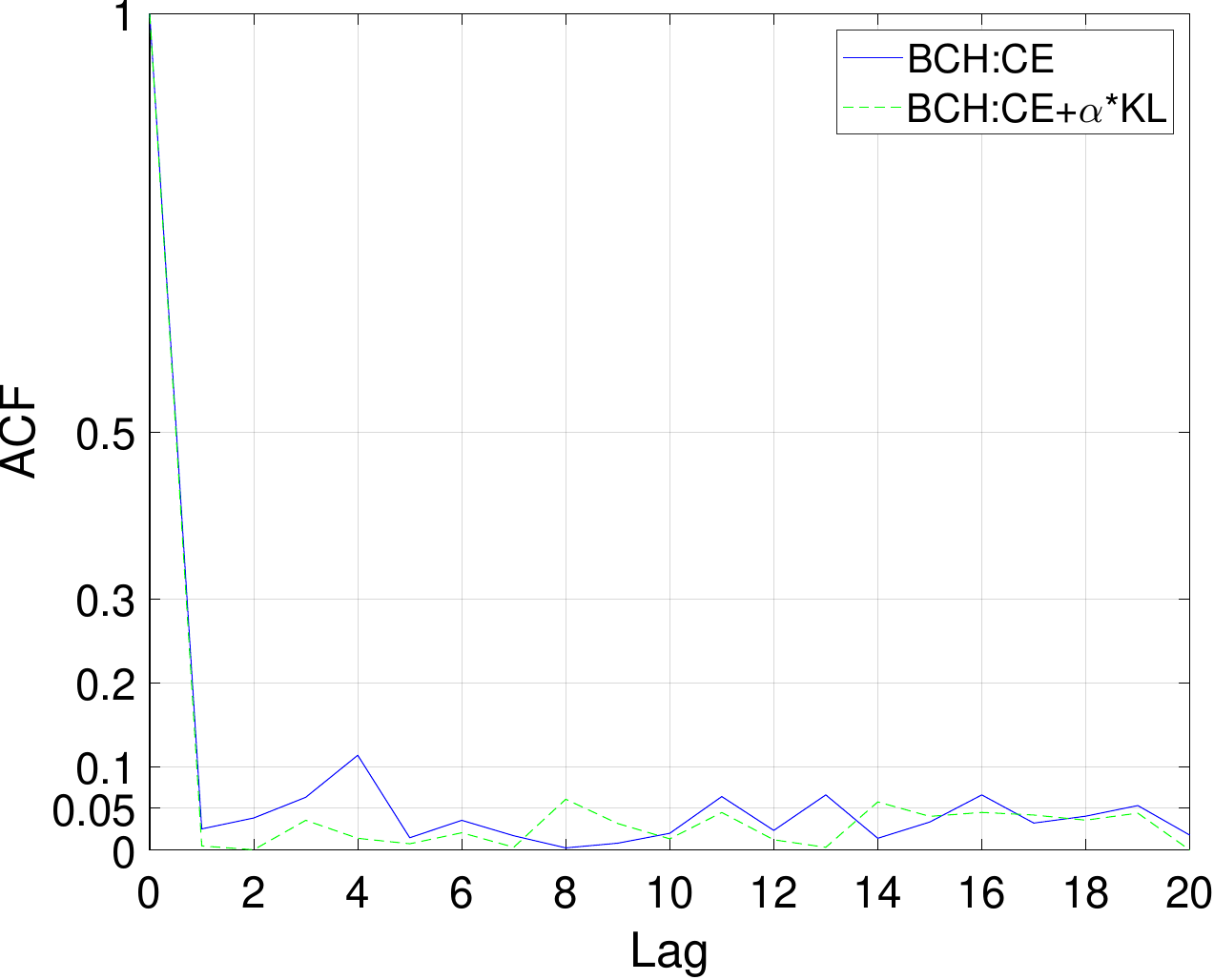}        
        \caption{\vspace{2em} ACF}
        \label{Fig7-2 BCH_KL_ACF}       
    \end{subfigure}
    \hfill
    \begin{subfigure}[b]{0.48\linewidth}
        \centering
       \includegraphics[width=\linewidth]{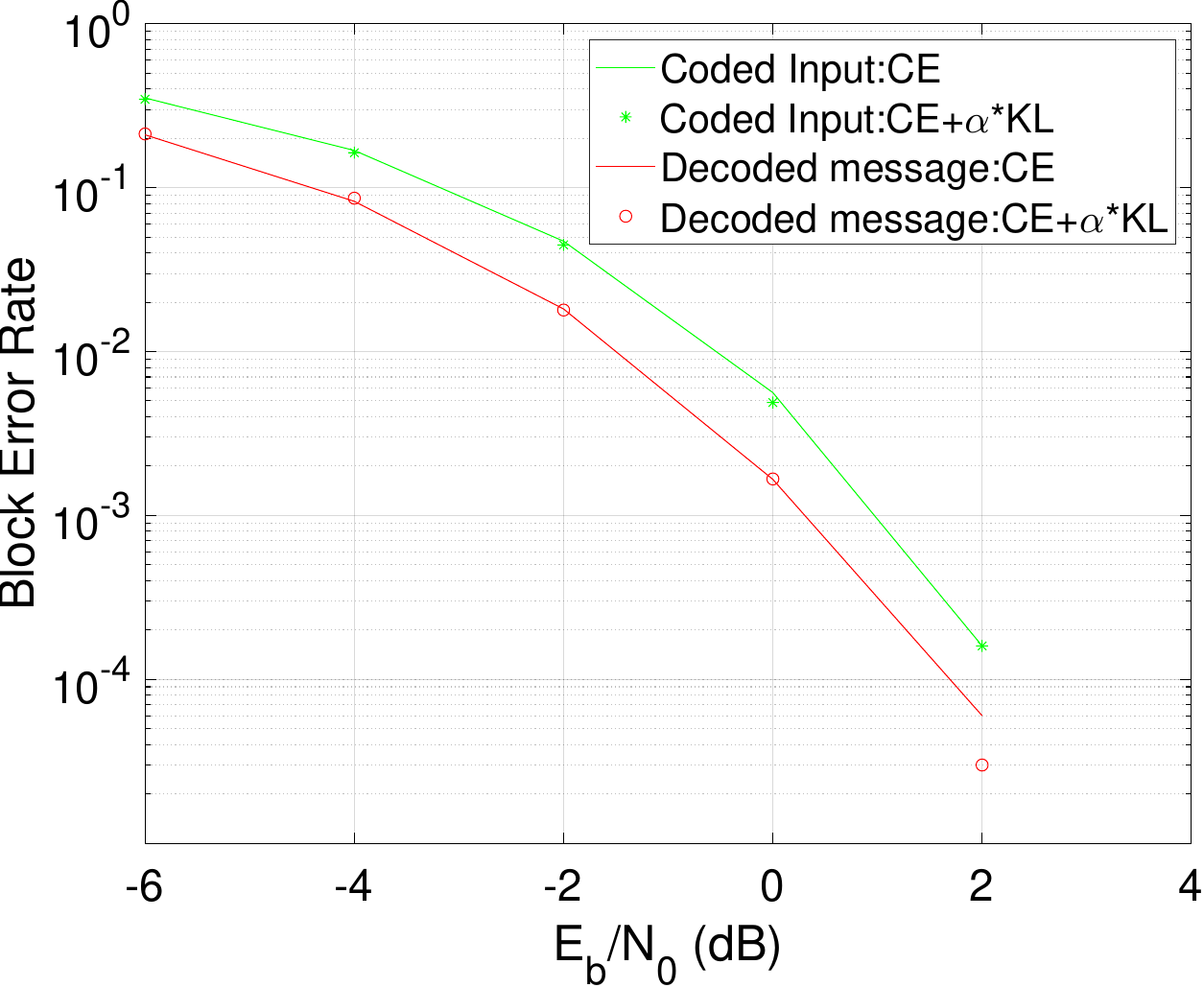}
        \caption{\vspace{2em} BLER}
        \label{Fig7-1 BCH_KL_BLER}
    \end{subfigure}
    \caption{Performance comparisons: In (a), ``BCH:CE'' refers to AE with BCH coded input trained with CE-only loss, and ``BCH:CE+$\alpha$*KL'' refers to AE with BCH coded input trained with combined CE and KL loss. In (b), AE with BCH coded input is trained with either CE-only or proposed loss.}
    \label{Fig7BCH_AE}
\end{figure}

\subsection{Performance of AE-based system on practical channel}
In this section, we demonstrate the AE‐based communication system in an over‐the‐air test. A single image of digit in MNIST dataset (Figure~\ref{Fig12-usrpimage}-(a)) is chosen and converted to a binary data and then segmented into 4‐bit blocks. Each block is utilized by the transmitter of a pre‐trained AE (\(k=4,n=20\)) on the fading channels, to produce AE symbols. A specially designed low‐ACF preamble is inserted before AE symbols to enable synchronization.

\begin{figure}[H]
    \centering
    \begin{subfigure}[b]{0.41\linewidth}
        \centering
         \includegraphics[width=\linewidth]{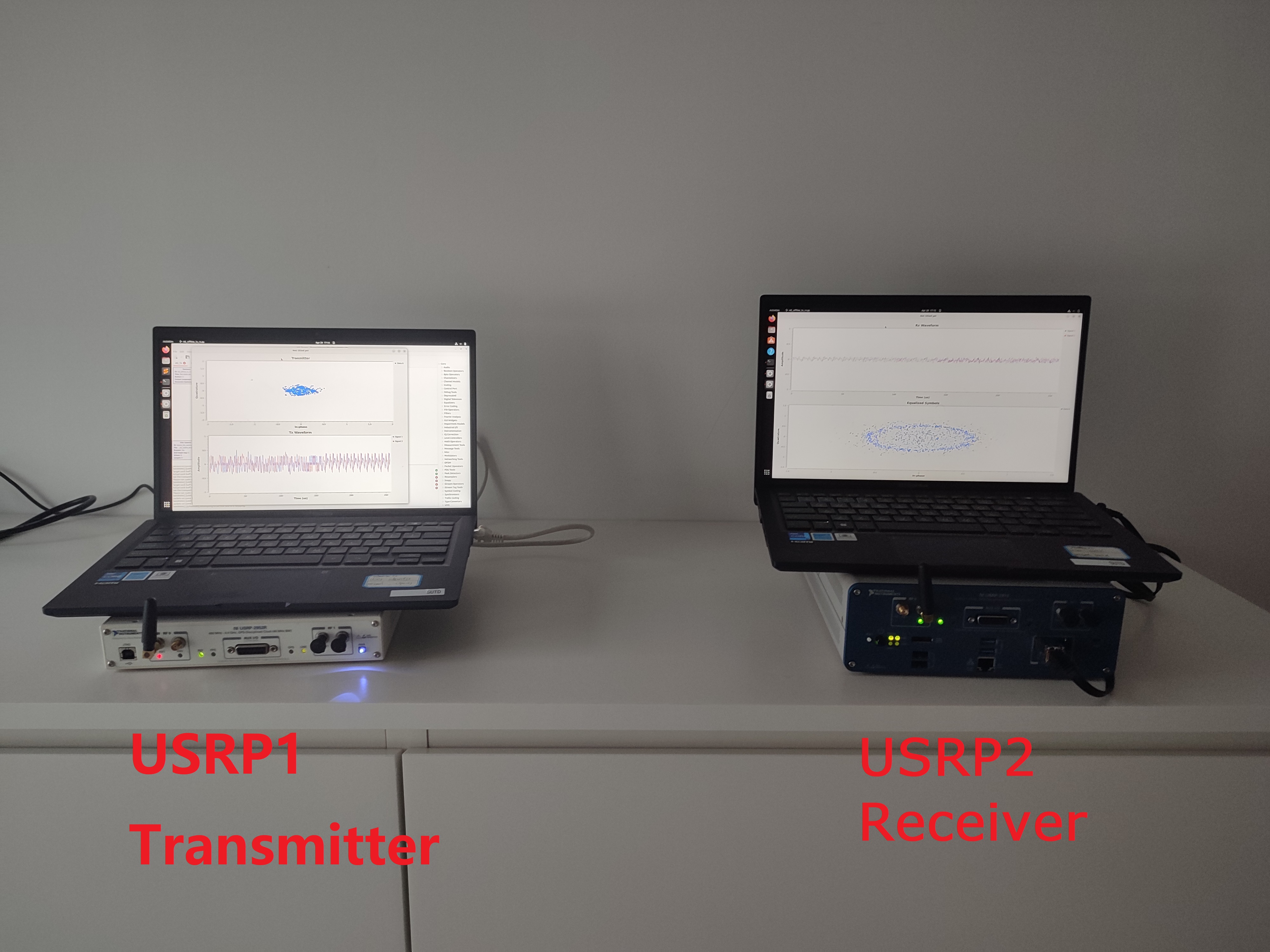}        
        \caption{\vspace{2em} USRP setting up}
        \label{Fig13 sourceImage}       
    \end{subfigure}
    \hfill
    \begin{subfigure}[b]{0.54\linewidth}
        \centering
       \includegraphics[width=\linewidth]{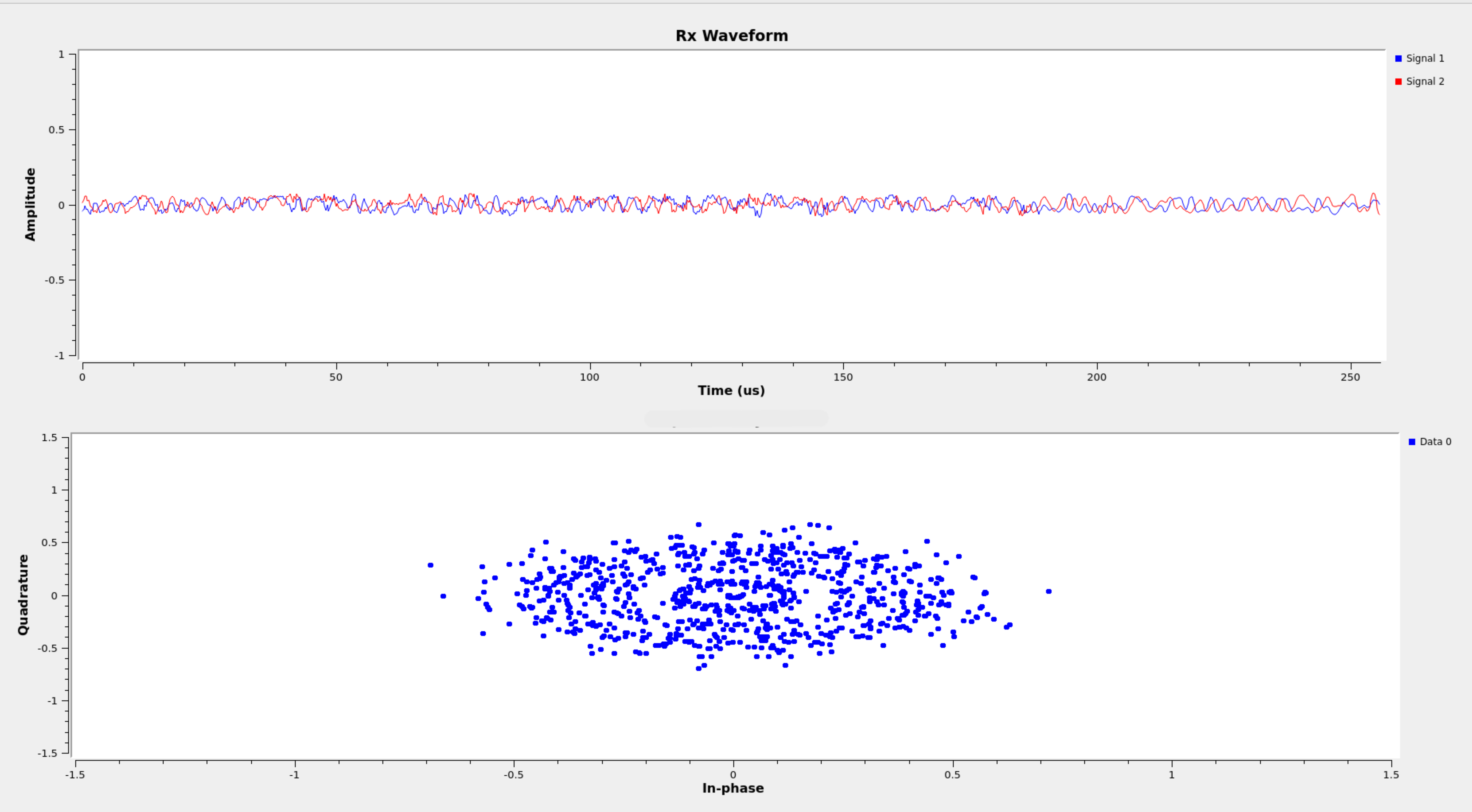}
        \caption{\vspace{2em} IQ Waveform and Constellation}
        \label{Fig13 RxImage}
    \end{subfigure}
    \caption{Practical communication using AE model}
    \label{Fig11-usrp}
\end{figure}

\begin{figure}[H]
    \centering
    \begin{subfigure}[b]{0.33\linewidth}
        \centering
         \includegraphics[width=\linewidth]{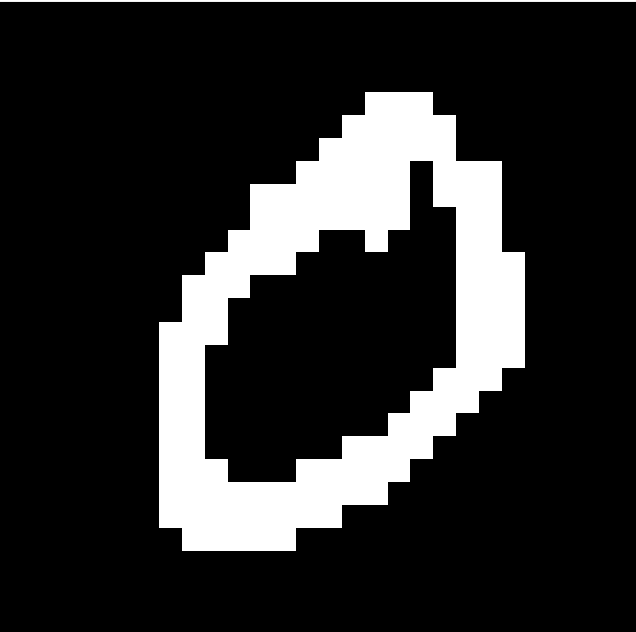}        
        \caption{\vspace{2em} Source Image}
        \label{Fig13 sourceImage}       
    \end{subfigure}
    \hfill
    \begin{subfigure}[b]{0.33\linewidth}
        \centering
       \includegraphics[width=\linewidth]{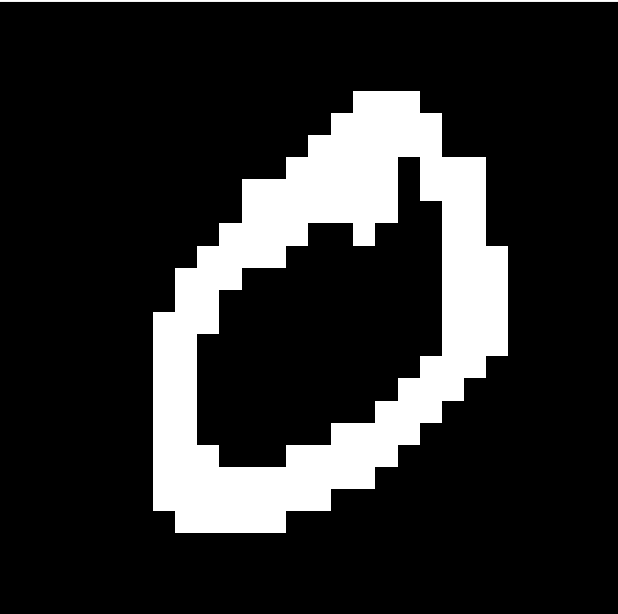}
        \caption{\vspace{2em} AE recovered Image}
        \label{Fig13 RxImage}
    \end{subfigure}
    \caption{Performance of AE on practical channel}
    \label{Fig12-usrpimage}
\end{figure}

Two Universal Software Radio Peripheral (USRP) units running GNU Radio communicate the preamble and AE‐generated symbols over the air, shown in Figure~\ref{Fig11-usrp}-(a). The resulting I/Q constellation (Figure~\ref{Fig11-usrp}-(b)) appears as a noise-like cloud with random amplitudes. At reception, we perform packet detection, frequency and timing synchronization, channel estimation and phase correction before feeding the synchronized symbols into the AE decoder. We observe the BLER of source blocks is zero and the reconstructed digit (Figure~\ref{Fig12-usrpimage}-(b)) perfectly matches the original (Figure~\ref{Fig12-usrpimage}-(a)). This over‐the‐air test demonstrates the viability of the AE‐based communication system in realistic wireless scenarios, achieving featureless wireless communication without conventional modulation and channel coding schemes. 

\section{Conclusion}\label{lbl:conclusion}

In this paper, we introduce two innovative approaches to enhancing AE-generated featureless signals for secure communication systems. First, by incorporating KL divergence into the CE loss function, we demonstrate that the AE-generated signal more closely resembles noise while maintaining the BLER performance. Secondly, our method of using binary inputs for AE encoded with conventional coding schemes significantly improves the detection performance of source message blocks. The AE with binary input is particularly suited for long message blocks with many bits, a scenario where the conventional AE with one-hot input becomes impractical or suffers from lower throughput. We observe that the use of coded binary inputs allows the AE to learn some characteristics of conventional coding schemes, resulting in lower BLER of coded blocks. Besides, the BLER of source message blocks is further reduced by the error-correction coding scheme. Moreover, when the AE with binary input is trained with the new loss, the AE signals also reveals the featureless property. Finally, we validate the AE-based communication system in the over-the-air communication using USRPs, demonstrating reliable transmission in a real wireless environment. These results underscore the promise of AE-driven designs for building secure and reliable wireless communication systems.



\bibliographystyle{IEEEtran}
\bibliography{FeaturelessBib}

\clearpage
\end{document}